\documentstyle[emulateapj,psfig]{article}

\begin{document}

\slugcomment{Published in {\bf Science in China}, Series A, 43 (2000), 439}

\title{Are there real orthogonal polarization modes in pulsar radio
emission?}

\author{R. X. Xu, G. J. Qiao\\
        CAS-PKU joint Beijing Astrophysical Center
        and Department of Astronomy, \\
        Peking University, Beijing 100781, China}

\altaffiltext{1}{email: rxxu@bac.pku.edu.cn}

\altaffiltext{2}{This work is supported by NSFC (No. 19803001),
by Doctoral Program Foungation of Institution of Higher
Education in China and by the Younth Foundation of PKU.}

\begin{abstract}

The Orthogonal Polarization Modes (OPM) have been reported observationally
(see e.g. [1]) and accepted widely by pulsar researchers (see e.g. [2,3]).
However, no acceptable theory can show the origin of the OPM, which becomes
a mystery in pulsar research field. Here a possible way to solve this mystery
is presented. We ask a question: Does there exist any real so-called OPM
in pulsar radiation? It is proposed in this paper that the `observed OPM'
in individual pulses could be the results of depolarization of pulsar
radiation and the observational uncertainties
originated from polarimeter in observation.
A possible way to check this idea is suggested. If the idea is verified,
the pulsar research would be influenced significantly in theory and in
observation.

\end{abstract}


\section{Introduction}

Pulsars are effective astrophysical laboratories for quantum theory and
gravitation theory. However, 
how to reproduce the observed radiation theoretically is still
one of the most essential challenges in pulsar study. It
is well known that the polarization observations are very important to
provide much information about pulsar physics, but there are still many
troubles in explaining the polarization data.

One of the difficulties in understanding pulsar polarization observations
is the polarization position angle jumps in mean (or integrated) pulses as
well as in individual pulses [1-3]. {\it For mean pulses}, it is generally
found that position angles would have discontinuities about 90$^o$ at some
longitudes where the linear polarization intensities are near zero (totally
depolarized). {\it For individual pulses}, the position angles would be
dispersed or have two $\sim 90^o$ separated distributions at some
observational longitude bins where the linear polarization percentages are
remarkably small. A famous example to display the polarization position
angle jumps in individual pulses is shown in Fig.1 for PSR B2020+28[1]. In
case `A' and `C', the two position angle distributions are clear, and the
linear polarization percentages are obviously low. In case `B', there is
only one position angle distribution, and the emission in each longitude
point is highly linearly polarized. It means that such orthogonally
distributed position angles are usually observed only in the part of the
profiles where the linear polarization is low [4]. A conclusion from the
observation is that, {\it both for mean profiles and for individual pulses,
the position angle jumps are related to low linear polarization (percentage)
at all time}.

Based on above observational facts, there should be two possibilities
logically: one is that `position angle jump' causes `low linear
polarization', another is that `low linear polarization' causes `position
angle jump'. Many authors believe, without justification, that the position
angle jumps should be attribute to the appearance of OPM [1-3], in-coherent
superposition of the OPM is the origin of depolarization. However, why the
another possibility is impossible? We investigate this possibility in this
paper.

Previously, Stinebring et al. [1] concluded that most of pulsar emission
occurs in one orthogonal mode or the other, which is called Orthogonal
Polarization Modes (OPM). At a given longitude, the plane of polarization
can be of two perpendicular or nearly perpendicular states. Which one can
operate was governed by some variability as yet not understood. The OPM can
explain many things: the sharp jumps of position angles, the depolarization
of linear as well as circular polarization due to the existence of both modes
at the same time [2]. However, there is no acceptable theory to reproduce such
orthogonal modes, which is called as `OPM problem'.

For mean pulses, it is suggested that the depolarization and position angle
jumps might be attribute to the relative longitude shifts of pulsar beams
[5,6]. Such kind of longitude shifts of pulsar beams is natural in the
inverse Compton scattering (ICS) model [7]. For individual pulses, many
authors believe that there are two orthogonal modes at a given longitude.
Nevertheless, there may be in fact two possibilities to produce such
orthogonal modes. One is that the emission for a given frequency is emitted
at different heights, and another is that there is an unknown emission
mechanism to produce orthogonal modes at a same emission point [8]. The first
possibility has been studied already [6, 7]. But for the second one, no
acceptable theory has been found to produce such orthogonal modes hitherto
known. Thus, the `OPM problem' still confuses the pulsar world.

We ask a question here: Are really such `orthogonal modes' the reason for
the `low linear polarization'?  Otherwise, might the `low linear
polarization' be responsible for the observed `orthogonal modes'? Many
authors believe that the reduction in the percentage of linear polarization
is caused by in-coherent superposition of the OPM. Contrary to the above
idea, another possibility (i.e., the `low linear polarization' is the reason
for producing the observed `position angle jumps') is suggested in this
paper. {\it Our analysis and simulations show that, when the linear
polarization percentages are low enough, the position angles would be
distributed in two areas separated nearly ninety degrees}. A suggestion to
check this idea is presented.

We show how `low linear polarization' causes so-called `position angle jumps'
in mean-pulses and in single-pulses in section 2 and 3, respectively. Some
troubles faced by OPM radiation mechanism are summed up in section 4. Finally,
conclusion and discussion are given in section 5.

\section{Position angle jumps in mean pulses: depolarization?}

Almost certainly, for observed mean-pulses, the smoothly changing position
angle curves will suddenly jump at some longitudes where the linear
polarization is highly depolarized. These facts can be understood under the
properties of Stokes parameters.
It could be verified mathematically that the position angle would jump
90$^o$ when the line of sight travels across a {\it singular point} [5, 6]
where the linear polarization intensity is zero.

The four Stokes parameters $\{I, Q, U, V\}$, from which one can obtain linear
polarization intensity $L=\sqrt{Q^2+U^2}$ and position angle $\chi$ (see
equ.(1) below), are functions of observational longitude $\phi$.
For the sake of simplicity, we let $V=0$, as the linear polarization is
focused on here. At
a singular point ($\phi=\phi_{\rm s}$), $L=0$ means
$
Q(\phi_{\rm s}) = 0,
U(\phi_{\rm s}) = 0.
$
Expanding $Q$ and $U$ near singular point, we come to
$$
Q(\phi_{\rm s} + \Delta) = {\partial Q\over \partial \phi}\Delta +
{1\over 2} {\partial^2 Q\over \partial \phi^2}\Delta^2 + {1\over 3!}
{\partial^3 Q\over \partial \phi^3}\Delta^3 + \cdot \cdot \cdot,
$$
$$
U(\phi_{\rm s} + \Delta) = {\partial U\over \partial \phi}\Delta +
{1\over 2} {\partial^2 U\over \partial \phi^2}\Delta^2 + {1\over 3!}
{\partial^3 U\over \partial \phi^3}\Delta^3 + \cdot \cdot \cdot.
$$
Assuming ${1\over q!} {\partial^q U\over \partial \phi^q}\Delta^q$ and
${1\over u!} {\partial^u U\over \partial \phi^u}\Delta^u$ are the lowest
non-zero
power terms of $Q$ and $U$, respectively, and $\nu = {\rm min}[q, u]$, one
could find that $\chi(\phi_{\rm s} + \Delta) - \chi(\phi_{\rm s} - \Delta)$
should be $\pm 90^o$ as long as $\nu$ is an odd number
\footnote{
One might easily obtain this conclusion by inspecting the position
angles in Q-U plane (two dimension Poincare sphere for $V=0$).
}
. It is very possible that $\nu=1$, thus, position angle {\it naturally}
jumps $90^o$ 
if $L=0$ [6]. Therefore, {\it the reason that position angle jumps in
integrated profiles might be why the beamed radiation is depolarized}.
Depolarization should be the cause of position angle jumps in mean-pulses.

There are many ways to cause depolarization. First of all, depolarization
may have an intrinsic origin. As emission beams are formed in
different heights, and each of them has different position angle,
depolarization must take place by incoherent superposition of such emission
beams. In the ICS model [7], different emission beams are formed in
different heights, hence, the retardation and aberration effects could make
the apparent emission beams be superposed incoherently [5,6]. Secondly,
depolarization might be originated from propagation process, such as the
scattering by interstellar medium or magnetospheric plasma [4], and the
propagating properties of different radiation modes in plasma. The third
way might be the result of observational effect. Since the Stokes parameters
are added from many frequency channels after de-dispersion, the emissions
in each frequency channel are incoherent superposed. Such kind of treatment
in observation should also depolarize the original radiation.

\section{Position angle jumps in single-pulses: observational
uncertainty?}

There are many factors to reduce the precision of observational results,
such as noises from the observational system and the sky background. Usually,
we put thresholds for the total intensity ($I$) and the linear polarization
intensity ($L$) in each longitude bin in order to exclude fake polarization
due to the observational error. For example, we select observational data
whose $I$ and $L$ are greater than 5 to 10 times of off-pulse rms. However,
as will be discussed in this section, some fake polarization data, which
may responsible for the observed `position angle jumps' in individual pulses
[5], do survive from such selection.

Some observational uncertainties can cause the observed position angles to
`jump' in individual pulses, such as the `error transference' (section 3.1),
the unequal rms of Stokes parameter $Q$ and $U$ (section 3.2), and the fake
linear polarization (section 3.3). All these uncertainties can bring wrong
polarization information.

\subsection{Position angle `jumps' in individual pulses due to the error
transference}

If $x$ is a random number, then the function $y=f(x)$ is also random. The
random distribution of $y$ is known as long as the distribution of $x$ is
given. For example, let the distribution function of $x$ as a gaussian
distribution, with the expectation value $x_0$. The distribution function
of $y$ depends on the function $f(x)$. If $f(x)$ is a monotonous
function near $x_0$, the distribution function of $y$ is approximately a
gaussian. Whereas, if $f(x)$ is a very complex function near
$x_0$, the distribution function of $y$ is also complicated.

It is known that the position angle $\chi$ is a function of the Stokes
parameters $Q$ and $U$
\footnote{
Usually $\chi = {1\over 2}\tan^{-1}{U\over Q}$, depending on the signs of
$Q$ and $U$. From
$
\left\{	\begin{array}{lll}
\sin 2\chi & = & U/L\\
\cos 2\chi & = & Q/L
\end{array}	\right.
$
, one can get a general expression for $\chi$ , where the value region of
$\chi$ is from $0^o$ to $180^o$.
}
$$
\chi={1\over 2} [{\rm sign}U \cos^{-1}{Q\over L} + \pi (1-{\rm sign}U)],
\eqno(1)
$$
which is `singular'(unusual) near $L=\sqrt{Q^2+U^2}=0$. Here ${\rm sign}U
= +1$ (${\rm sign}U = -1$) if ${\rm sign}U > 0$ (${\rm sign}U < 0$). This
property of singularity of $\chi$ would cause two error distribution peaks
(see Fig.2), which will be discussed later in section 3.2. In a word, the
observational uncertainty (error) of $Q$ and $U$ could bring an error
distribution of $\chi$ in two regions with 90$^o$ separation by the error
transference effect. Such observational results might be mistakenly
considered as position angle `jumps' in a real beamed radiation.

\subsection{Position angle jumps due to unequal errors of $Q$ and $U$}

Usually, the rms of the Stokes parameter $Q$ and that of $U$ are not equal
for an astronomical polarimetry, i.e., $\sigma_{\rm Q} \neq \sigma_{\rm U}$.
The difference between $\sigma_{\rm Q}$ and $\sigma_{\rm U}$ can be as large
as several percentages in observations. The observed linear polarization
position angles could `jump' during different observing time as long as
$\sigma_{\rm Q} \neq \sigma_{\rm U}$. As demonstrated in Fig.3, we see that
$\chi_B - \chi_A$ is about $0^o$, and $\chi_B - \chi_C$ is about $90^o$.

The reason that $\sigma_{\rm Q}\neq \sigma_{\rm U}$ might be diversity. For
example, for some polarimetry, the Stokes parameters $Q=S_0-S_{90}$ and
$U=S_{45}-S_{135}$ are computed from the hybrid networks' out-put signals.
Here $S_0$ and $S_{90}$ are the observed intensity from two orthogonal
dipole antenna, $S_{45}$ and $S_{135}$ are the intensity received from a
system which has been rotated 45 degrees.
For a dipole antenna, $S_0$ and $S_{90}$
are obtained directly. However, the $S_{45}$ and $S_{135}$ are yielded
through a turnstile junction where the phase mis-alignment can make the
rms of $S_{45}$ (and $S_{135}$) larger than the rms of $S_0$ (and $S_{90}$).
So, the rms of $Q$ and that of $U$ can not be equal because of the imperfection
of the turnstile junction.

Simulations of this kind of polarimetry are given in Fig. 4
(see the Appendix), from which we see that

\begin{itemize}

\item The observed percentages of polarization $\Pi'$ are much greater than
the true percentage $\Pi$, if $\Pi$ is small enough. Observed linear
polarization may be larger than that of the true value.

\item The position angle `jumps' takes place when linear polarization
percentage $\Pi_l \leq 0.1\%$. When $\Pi_l \geq 1\%$, there is few
possibilities to make position angle jump.

\end{itemize}

Because the observational uncertainty is of random, position angle jumps
that come of this kind of errors discussed above can be avoided by more time
observation. It is almost impossible that position angle jump due to
observational uncertainty appears in integrated profiles.

\subsection{The observational noise responsible for fake polarization}

For a telescope with an effective detection area $A$, a frequency bandwidth
$\delta \nu$, a time constant $\tau$, and a systematical noise temperature
$T_{\rm sys}$, then the systematical noise flux $S_{\rm sys}$ is
$
S_{\rm sys} = k T_{\rm sys}/A
$,
where $k=1.38\times 10^{-23}{\rm JK}^{-1}$ is the Boltzmann's constant, and
the off-pulse rms $\sigma_{\rm off}$ is $$
\sigma_{\rm off} = {S_{\rm sys} \over \sqrt{\delta \nu \ \tau}}.
$$
Whereas there is a signal with intensity flux $S_{\rm i}$, the rms of the
signal flux is $\sigma_{\rm on}$ (the on-pulse rms)
$$
\sigma_{\rm on} = {S_{\rm sys} + S_i \over \sqrt{\delta \nu \ \tau}},
$$
where $S_{\rm i}$ can be one of the out-put signals from the hybrid networks,
such as $S_0$, $S_{90}$, $S_{45}$, $S_{135}$, $S_R$, $S_L$, which are
correspondent to the linearly polarized components of the input signals at
position angle $0^o$, $90^o$, $45^o$, and $135^o$, and to the right-hand
and left-hand circularly polarized components.

Because $S_{\rm i} = (S_{\rm sys} + S_{\rm i}) - S_{\rm sys}$, the rms of
$S_{\rm i}$ is
$
\sigma_{\rm i} = \sqrt{\sigma_{\rm off}^2 + \sigma_{\rm on}^2}
$.
If we assume that $S_{\rm sys}$ is accurate enough (i.e., we have enough
time to measure $S_{\rm sys}$), then
$
\sigma_{\rm i} \approx \sigma_{\rm on}
$,
which will be used in the following discussion.

If we let $A = 40^2 \pi$ square meters, $\delta \nu = 10$MHz, $\tau=0.3$ms,
and $T_{\rm sys} = 40$K, then $\sigma_{\rm off}$ is about 0.2 Jy. For
$S_i=30$Jy[9], then $\sigma_i=0.8 {\rm Jy}$ which is more than three times
that of $\sigma_{\rm off}$, and the fake linear polarization percentage
could be as large as $3\%$. Thus, some data, whose linear polarization
is originated from such uncertainty, {\it can also exceed} the observational
threshold level with an un-negligible possibility.

\section{ Are real orthogonal polarization modes in pulsar radiation?}

We might be in a dilemma if there are really so-called orthogonal
polarization modes in pulsars' beams. First, how does the OPM radiative
mechanism produce? No reasonable theory has been appeared in literature.
Furthermore, the two orthogonal modes should be in-coherent, which makes
the OPM more difficult to be set up. The suggestion for observed `orthogonal
modes' in mean-pulses by Xu at al. [6] is not a real one. (In their
calculations, two components are emitted in-coherently from different
regions.) Furthermore, If the two modes are coherent, the total radiation
is elliptically polarized, thus no position angle `jump' appears.

Secondly, why haven't we seen that single-pulses are highly polarized, but
the position angle distribution is till separated by 90$^o$ (like Fig.5)?
Individual pulses, which could be highly polarized, are generally
conjectured to be from single radiation elements. Since two orthogonal
modes are incoherent, a radiation element might emit only one of the OPR
modes at one time. Therefore, it is possible to observe some highly linearly
polarized individual pulses, while the position angles of which are 90$^o$
separated. Unfortunately, observation result similar to that of Fig.5 has
{\it never} been found.

Thirdly, how to explain the non-orthogonal separation of position angles in
the regime of OPM? Non-orthogonal emission modes have proverbially been
found in observation [1]. These facts are rigorous for anyone to theorize
OPM models.

\section{Conclusion and discussion}

From analysis above some conclusions and discussions are reached:

1. Another possible way to solve the problem of position angle `{\it jumps}'
in pulsars' beamed radio emission was proposed. There might be {\it no real}
`orthogonal polarization modes' in the emission at all.

2. Position angle jumps due to the observational uncertainties could appear
in observed individual pulses when the linear polarization {\it percentages}
are small (not only the linear polarization intensity to be small). At least
part of the observed position angle jumps in individual pulses and
mean-pulses can be explained by depolarization and observational
uncertainty.

For a real pulsar, we must put together these two possible factors to
investigate the position angle variation in the individual pulses as well
as in the integrated pulses. For example, observational uncertainty might
be the main reason of position angle separation near point `A' in Fig.1.
Nevertheless, near point `C', orthogonal and non-orthogonal separations are
clear, which might be the result of the relative longitude
shifts of pulsar beams
[6] and the observational uncertainties.

Rathnsree \& Rankin[10] pointed out that, for PSR B1929+10, lower degree
of polarization is seen simultaneously with the presence of `orthogonal' modes
whereas the polarized power is not seen to be highly correlated with the
position angle flip. Also, they have got dynamic pictures of the orthogonal
polarization mode changes for PSR B2110+27 at 430 MHz, and they found the
transition from the dominant mode to the other orthogonal one
and back are {\it rapid}.
Most of the transition is achieved over time scales of a individual
period, and the change of modes does not seem to be any periodicity in time
evolution (like a {\it stochastic} process). All this observational facts 
have the properties of observational uncertainties discussed in section 3.

For PSR B0525+21[1] at 1404 MHz, the position angle sweep is S-shaped
in the averaged profile. The polarization position angles in individual
pulses are also an `S' shape distribution, but two weak patches of
`orthogonal modes' on the outside edges of the profile where
more individual pulses have very small percentages of linear polarization.
These two patches' appearance should come of the observational
uncertainty according to our analysis.
This statement {\it can be checked} by
future expriment of observation.

3. From simulations, we see that the jumped position angles are distributed
near $45^o$ and $135^o$. In fact, there are observational data which does
show that the position angles distribute near $45^o$ and $135^o$ in the
scatter plots, such as Fig.26, Fig.37 in Stinebring et al. [1]. If the rms
of $U$ less than that of $Q$, the jumped position angles should be near $0^o$
and $90^o$, like PSR 0525+21 (Fig.2 in [1]) in observations. There are
observational data where the jumped position angles are not distributed near
$0^o$, $45^o$, $90^o$, and $135^o$, which could be intrinsic in polarimetry
or resultant from the longitude shift of beam phases [6].

4. The idea suggested in this paper can be checked experimentally. We can
input the polarimetry a simulated lower polarized and pulsed signal to see
if two $90^o$ separated position angle distribution can appear in the output.
If such distribution can also be obtained, the OPM in pulsar emission should
be doubted.

5. If OPM does not exist in pulsar radio emission, we should develop our
instruments to avoid observational uncertainties.

\acknowledgements{We thank our pulsar group for discussions.}

\begin{center}
\Large{{\bf Appendix} A Simulation}
\end{center}

For the kind of polarimetry discussed in section 3, let's study a partially
polarized wave, with the total intensity $I$, the un-polarized intensity
$I^{unp}$, the percentage of linear polarization $\Pi_l$, the percentage
of circular polarization $\Pi_c$, and the position angle $\chi$. If we
measure this wave by a telescope with an effective area $A$, systematic noise
temperature $S_{\it sys}$, a bandwidth $\delta\nu$, a time constant $\tau$,
the angle between two the dipole antenna (parasitism polarization) $\alpha$
(whose expectation value is $\pi\over 2$, $\delta\alpha = \alpha -{\pi\over
2}$, the rms of $\delta\alpha$ is $\sigma_{\alpha}$), and the phase
misalignment $\delta\phi$ (the rms of $\delta\phi$ is $\sigma_{\phi}$),
then the six intensity for the Stokes parameters can be deduced as
$$
\begin{array}{lll}
S_0 & = & {1\over 2}X^2 + {1\over 2}I^{unp},\\
S_{90} & = & {1 \over 2}(Y^2 \cos^2 \delta\alpha +\\
& & X^2 \sin^2 \delta\alpha +2XY\cos\delta\sin\delta\alpha\cos\delta\alpha) +\\
& & {1\over 2}I^{unp},\\
s_{45} & = & {1\over 4} [X^2 + Y^2\cos^2\delta\alpha +\\
& & X^2\sin^2\delta\alpha +\\
& & 2XY\cos(\delta-\delta\phi)\cos\delta\alpha + \\
& & 2XY\sin\delta\alpha\cos\delta\alpha\cos\delta +\\
& & 2X^2\sin\delta\alpha\cos\delta\phi] + {1\over 2}I^{unp},\\
S_{135} & = & {1\over 4} [X^2 + Y^2\cos^2\delta\alpha +\\
& & X^2\sin^2\delta\alpha +\\
& & 2XY\cos(\delta-\delta\phi-\pi)\cos\delta\alpha+\\
& & 2XY\sin\delta\alpha\cos\delta\alpha\cos\delta +\\
& & 2X^2\sin\delta\alpha\cos(\delta\phi+\pi)] + {1\over 2}I^{unp},\\
S_R & = &{1\over 4} [X^2 + Y^2\cos^2\delta\alpha +\\
& & X^2\sin^2\delta\alpha +\\
& & 2XY\cos(\delta+\delta\phi+{3\over 2}\pi)\cos\delta\alpha+\\
& & 2XY\sin\delta\alpha\cos\delta\alpha\cos\delta +\\
& & 2X^2\sin\delta\alpha\cos(\delta\phi+{3\over 2}\pi)]+{1\over
2}I^{unp},\\
S_L & = &{1\over 4} [X^2 + Y^2\cos^2\delta\alpha + X^2\sin^2\delta\alpha
+\\
& & 2XY\cos(\delta+\delta\phi+{\pi\over 2})\cos\delta\alpha+\\
& & 2XY\sin\delta\alpha\cos\delta\alpha\cos\delta +\\
& & 2X^2\sin\delta\alpha\cos(\delta\phi+{\pi\over 2})]+{1\over
2}I^{unp},
\end{array}
$$
here,
$$
\begin{array}{lll}
\delta & = & \tan^{-1}{\Pi_c\over \Pi_l\sin2\chi},\\ X & = & \sqrt{I} \times
\sqrt{\sqrt{\Pi^2_l + \Pi^2_c} + \Pi_l\cos2\chi},\\
Y & = & \sqrt{I} \times \sqrt{\sqrt{\Pi^2_l + \Pi^2_c} - \Pi_l\cos2\chi}.\\
\end{array}
$$
The observed Stokes parameters should be $$
\begin{array}{lll}
I' & = & S_0 + S_{90},\\
Q' & = & S_0 - S_{90},\\
U' & = & 2S_{45} - I',\\
V' & = & 2S_R - I'.\\
\end{array}
$$
So that, the observed linear polarization intensity $L'$, the observed
percentages of linear polarization $\Pi'_l$, the observed percentages of
circular polarization $\Pi'_c$, and the observed linear polarization
position angle $\chi'$ would be $$
\begin{array}{lll}
L' & = & \sqrt{Q'^2+U'^2},\\
\Pi'_l & = & {L'\over I'},\\
\Pi'_c & = & {V'\over I'},\\
\chi' & = & {1\over 2} [{\rm sign}U' \cos^{-1}{Q'\over L'} + \pi (1-{\rm
sign}U')].
\end{array}
$$

Considering this kind of observational uncertainty, we have obtained some
simulation results to show the position angle `jumps' in individual pulse
observations. One of the simulations is shown in Fig.4, where we have chosen
$$
\begin{array}{lll}
I & = & 50\;{\rm \;Jy},\\
\sigma_{\alpha} & = & 5^o,\\
\sigma_{\phi} & = & 5^o,\\
A & = & \pi\;40^2 \;{\rm \;square\;meters},\\ S_{sys} & = & 40\;{\rm \;K},\\
\delta\nu & = & 10\;{\rm \;MHz},\\
\tau & = & 0.3\;{\rm \;ms},\\
\chi & = & 45^o,\\
\Pi_l & = & \Pi_c=\Pi=0.01\%.
\end{array}
$$
The scatter plots in Fig.4 are resemble to observations, especially the
$L'-\chi'$ plot, which is similar to the observed linear polarization versus
position angle scatter plots for position angle jumps at a fixed longitude.
Based on this simulation and other simulations for different parameters,
we found that the position angles `jump' if $\Pi_l \leq 0.1\%$, whereas,
there is few possibilities of position angle jump if $\Pi_l \geq 1\%$.

{\bf Figures:}

\vspace{0.2cm}
\centerline{}
\centerline{\psfig{file=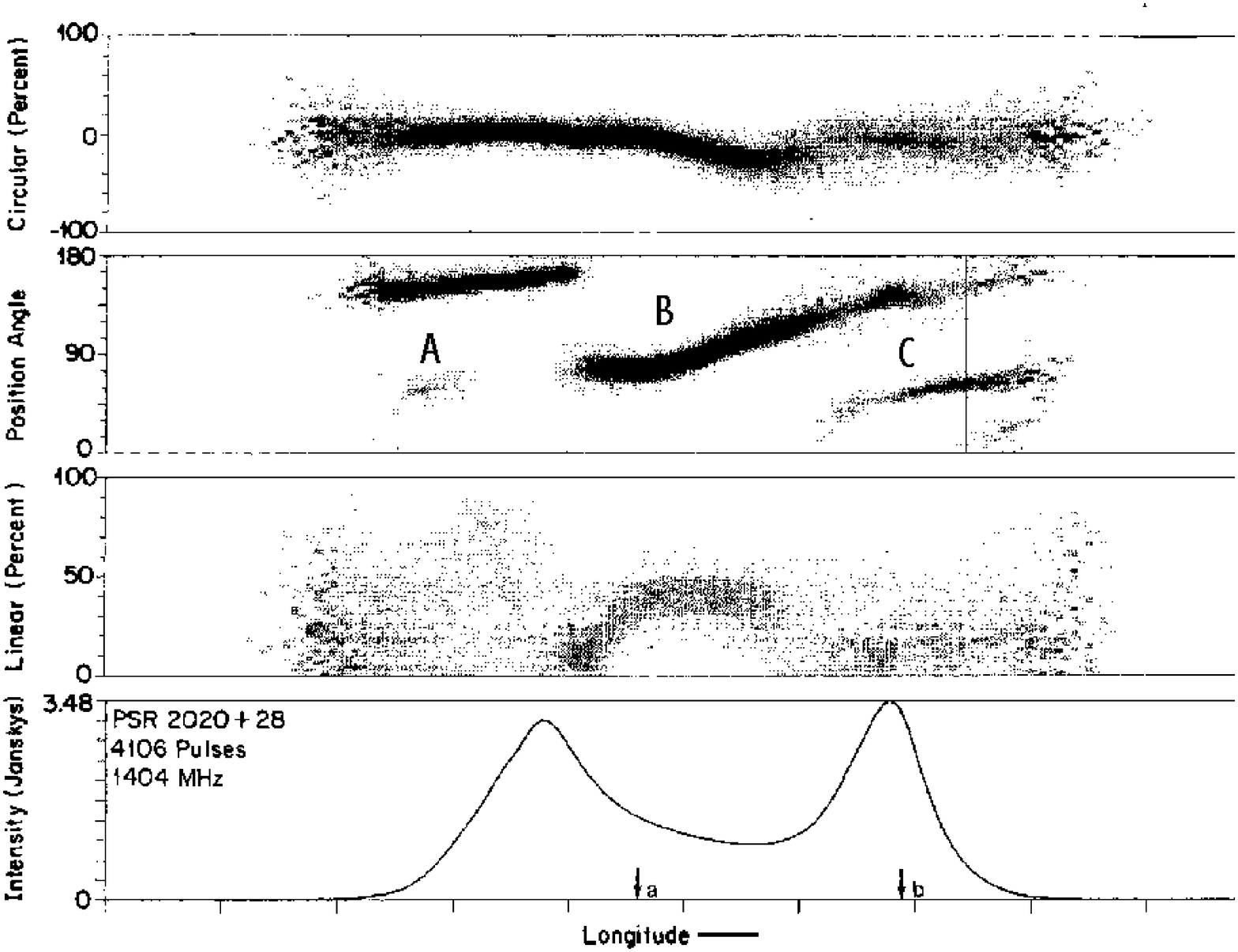,width=20cm,height=15cm}}
\figcaption{
Polarization distribution of PSR B202+28. The top plot shows the
position angle distribution; the middle plot shows the linear polarization
percentage of individual pulses; the lowest plot gives the integral pulse
profile. The observation is done by Stinebring et al. [1].
\label{Fig.1}  }

\centerline{}
\centerline{\psfig{file=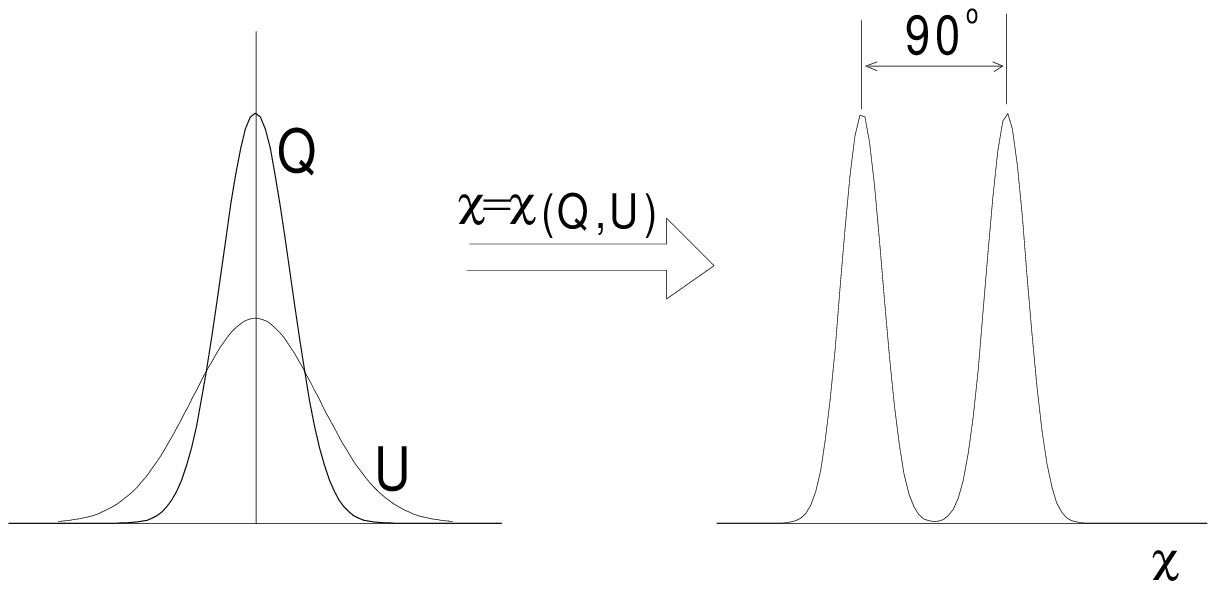,width=15cm,height=10cm}}
\figcaption{
A sketch picture for the possibility of the linear polarization
position angle `jump' due to the error transferring from $Q$ and $U$.
\label{Fig.2}  }

\centerline{}
\centerline{\psfig{file=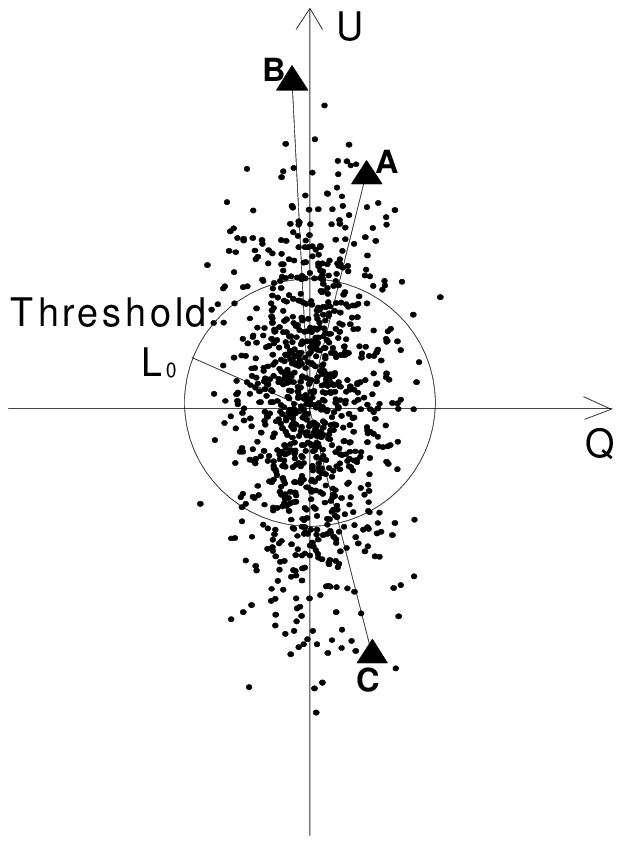,width=10cm,height=12cm}}
\figcaption{
A demonstration of position angle `jumps' which come from
observational uncertainty. Position angles $\chi_A$, $\chi_B$ and $\chi_C$
are for points A, B, and C, respectively.
\label{Fig.3}  }

\centerline{}
\centerline{\psfig{file=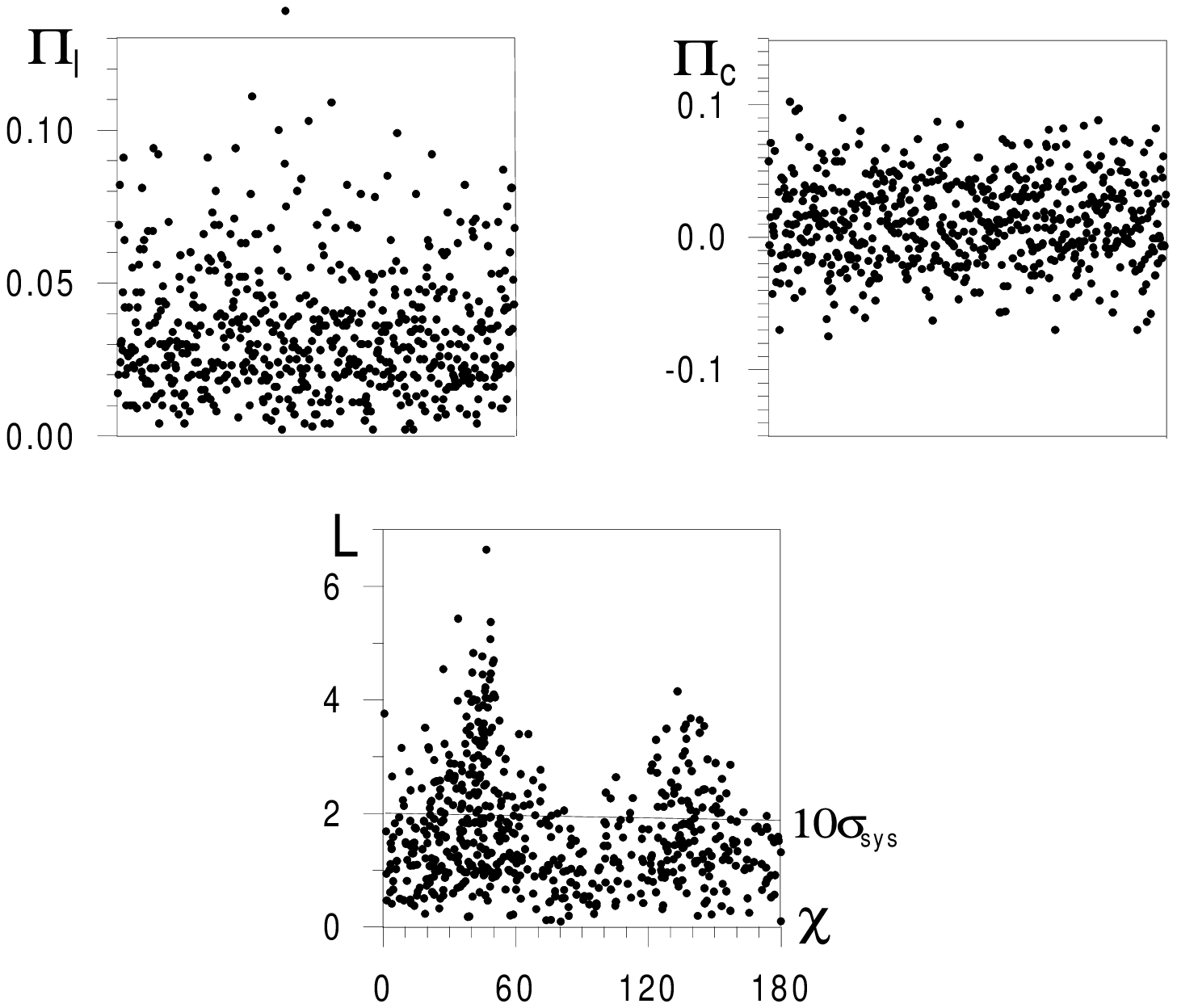,width=20cm,height=14cm}}
\figcaption{
The simulated scatter plots of the observed polarization degrees of
linear polarization $\Pi'_l$, and circular polarization $\Pi'_c$. The lower
one is the $L'-\chi'$ (linear polarization intensity vs. the position angle)
plot, where the solid horizontal line shows a possible threshold level for 
linear polarization. The parameters in the simulation are given in the text.
\label{Fig.4}  }

\centerline{}
\centerline{\psfig{file=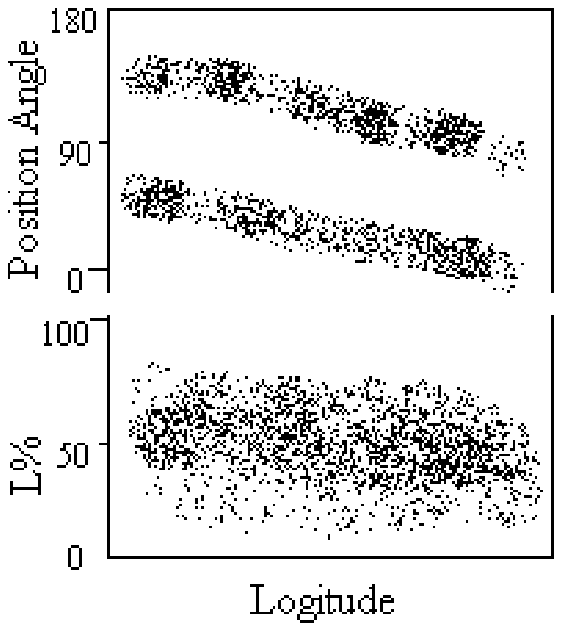,width=7cm,height=7cm}}
\figcaption{
A possible observational result predicted by OPM models.
In the figure, the lowest points present the linear polarization
persentages of individual pulses; the upper two distributions are
for the position angles.
\label{Fig.5}  }

\end{document}